\documentclass[final,pre,nofootinbib,twocolumn,showpacs,showkeys,groupaddress,preprintnumbers,floatfix]{revtex4-1}

\usepackage{dynlearn}

\renewcommand{\FCS}{\ensuremath{\CausalState_0^+}}
\renewcommand{\RCS}{\ensuremath{\CausalState_0^-}}

\newcommand{\funcset}[1]{\ensuremath{\bf{F}_{#1}}}
\newcommand{\mcset}[2]{\ensuremath{\bf{M}_{#1 #2}}}
\newcommand{\ssset}[2]{\ensuremath{\bf{S}_{#1 \rightarrow #2}}}

\begin{document}

\def\ourTitle{%
  Trimming the Independent Fat:\\
  Sufficient Statistics, Mutual Information, and\\
  Predictability from Effective Channel States
}

\def\ourAbstract{%
One of the most fundamental questions one can ask about a pair of random
variables $X$ and $Y$ is the value of their mutual information. Unfortunately,
this task is often stymied by the extremely large dimension of the variables.
We might hope to replace each variable by a lower-dimensional representation
that preserves the relationship with the other variable. The theoretically
ideal implementation is the use of \emph{minimal sufficient statistics}, where
it is well-known that either $X$ or $Y$ can be replaced by their minimal
sufficient statistic about the other while preserving the mutual information.
While intuitively reasonable, it is not obvious or straightforward that both
variables can be replaced simultaneously. We demonstrate that this is in fact
possible: the information $X$'s minimal sufficient statistic preserves about
$Y$ is exactly the information that $Y$'s minimal sufficient statistic
preserves about $X$.
As an important corollary, we consider the case where one variable is a
stochastic process' past and the other its future and the present is viewed as
a memoryful channel. In this case, the mutual information is the channel
transmission rate between the channel's effective states. That is, the
past-future mutual information (the excess entropy) is the amount of
information about the future that can be predicted using the past. Translating
our result about minimal sufficient statistics, this is equivalent to the
mutual information between the forward- and reverse-time causal states of
computational mechanics. We close by discussing multivariate extensions to this
use of minimal sufficient statistics.
}

\def\ourKeywords{%
  information theory, sufficient statistics, mutual information, dimensionality reduction, stochastic process, \texorpdfstring{\eM}{epsilon-machine}, causal states, transmission rate
}

\hypersetup{
  pdfauthor={Ryan G. James},
  pdftitle={\ourTitle},
  pdfsubject={\ourAbstract},
  pdfkeywords={\ourKeywords},
  pdfproducer={},
  pdfcreator={}
}

\author{Ryan G. James}
\email{rgjames@ucdavis.edu}

\author{John R. Mahoney}
\email{jrmahoney@ucdavis.edu}

\author{James P. Crutchfield}
\email{chaos@ucdavis.edu}

\affiliation{Complexity Sciences Center and Physics Department,
University of California at Davis, One Shields Avenue, Davis, CA 95616}

\date{\today}
\bibliographystyle{unsrt}

\title{\ourTitle}

\begin{abstract}
\ourAbstract
\end{abstract}



\preprint{\sfiwp{17-01-XXX}}
\preprint{\arxiv{1701.XXXX}}

\title{\ourTitle}
\date{\today}
\maketitle

\setstretch{1.1}

\section{Introduction}
\label{sec:introduction}

How do we elucidate dependencies between variables? This is one of the major
challenges facing today's data-rich sciences, a task often stymied by the curse
of dimensionality. One approach to circumventing the curse is to reduce each
variable while still preserving its relationships with others. The maximal
reduction---the minimal sufficient statistic---is known to work for a single
variable at a time~\cite{Cove06a}. In the multivariate setting, though, it is
not straightforward to demonstrate that, as intuition might suggest, all
variables can be simultaneously replaced by their minimal sufficient
statistics. Here, we prove that this is indeed the case in the two and three
variable settings.

The need for sufficient statistics arises in many arenas. Consider, for
example, the dynamics of a complex system. Any evolving physical system can be
viewed as a communication channel that transmits (information about) its past
to its future through its present~\cite{Crut08a}. Shannon information theory
\cite{Cove06a} tells us that we can monitor the amount of information being
transmitted through the present by the past-future mutual information---the
\emph{excess entropy}~\cite{Crut01a}. However, this excess entropy can rarely
be calculated from past and future sequence statistics, since the sequences are
semi-infinite. This makes calculating the excess entropy an ideal candidate for
using sufficient statistics. The latter take the form of either a process'
prescient states or its causal states \cite{Crut12a}. Though known for some
time~\cite{Crut08a}, a detailed proof of this relationship was rather involved,
as laid out in Ref.~\cite{Crut10d}.

The proof of our primary result turns on analyzing the information-theoretic
relationships among four random variables $W$, $X$, $Y$, and $Z$. All possible
informational relationships---in terms of Shannon multivariate information
measures---are illustrated in the information diagram~\cite{Reza61a,Yeun08a}
(I-diagram) of Fig.\nobreakspace \ref {fig:XWZY_idiagram}. This Venn-like diagram decomposes the
entropy $\H{X,Y,Z,W}$ of the joint random variable $(X,Y,Z,W)$ into a number of
\emph{atoms}---informational units that cannot be further decomposed using the
variables at hand. For example, take the region labeled $f$ in
Fig.\nobreakspace \ref {fig:XWZY_idiagram}; this region is the conditional entropy $\H{X \mid Y,
Z, W}$. Similarly, one has the four-variable mutual information $k =
\I{X:Y:Z:W}$ and the condition mutual information $d = \I{W:Z \mid X, Y}$. The
 analogy with set theory, while helpful, must be handled with care:
Shannon informations form a \emph{signed} measure. Any atom quantifying
the information shared among at least three variables can be negative.
In the context of our example, Fig.\nobreakspace \ref {fig:XWZY_idiagram}, atoms $g$, $h$, $m$, $n$, and $k$ can be negative.
Negative information has led to a great deal of investigation; see, for
example, Refs.~\cite{Jame11a,Will10a}.

Here we are interested in what happens when $W$ is a sufficient
statistic of $X$ about $Y$ \emph{and} $Z$ is a sufficient
statistic of $Y$ about $X$~\cite{Cove06a}. We denote this $W = X \mss Y$ and $Z = Y \mss X$.
The resulting (reduced) I-diagram provides a
useful and parsimonious view of the relations among the four variables.
In particular, it leads us to the main conclusion that each variable can be \emph{simultaneously} reduced to its sufficient statistic while maintaining the mutual informations. Our development
proceeds as follows: Section\nobreakspace \ref {sec:sufficient_statistics} defines sufficient
statistics and utilizes two of their properties to reduce the informational
relationships among the variables. Section\nobreakspace \ref {sec:channels} discusses how this
result applies to stochastic processes as communication channels. Section\nobreakspace \ref {sec:multivariate_extensions}
extends our results to the three variable case and makes a conjecture about broader applicability. Finally,
Section\nobreakspace \ref {sec:conclusion} outlines further directions and applications.

\section{Sufficient Statistics}
\label{sec:sufficient_statistics}

A \emph{statistic} is a function $f(\bullet)$ of random variable
samples~\cite{Cove06a}. Let \funcset{X} denote the set of all functions of a
random variable $X$. These functions are also random variables. Given variables $X$ and $Y$, a variable $V$ forms a \emph{Markov chain} $X - V - Y$ if $p(x, y) = \sum_{v} p(x)p(v|x)p(y|v)$.
Let \mcset{X}{Y} denote the set of all variables that form a Markov chain with $X$ and $Y$.
A \emph{sufficient statistic} of $X$ about $Y$ is an element of
$\ssset{X}{Y} = \funcset{X} \cap \mcset{X}{Y}$.\footnote{Our definition here is
equivalent to that provided in, \eg, Ref.~\cite{Cove06a}, but in a form
that more directly emphasizes the properties we exploit over the next two subsections.}
The \emph{minimal sufficient statistic} $X \mss Y$ of $X$ about $Y$ is the minimal-entropy sufficient statistic:
\begin{align}
  X \mss Y = \argmin_{V} \left\{ \H{V} \mid V \in \ssset{X}{Y} \right\}
  ~.
\end{align}
It is unique up to isomorphism~\cite{Shal98a}.

The minimal sufficient statistic can be directly constructed from variables $X$
and $Y$. Consider the function $f(\bullet)$ mapping $x$ to the conditional
distribution $p(Y | X = x)$; then $X \mss Y \sim
f(X)$~\cite{kamath2010new,wolf2004zero}. Put more colloquially, $X \mss Y$
aggregates the outcomes $x$ that induce the same conditional distribution $p(Y | X = x)$. This is an equivalence class over $X$, where the probability of each class is the sum of the probabilities of the outcomes contained in that class.

\subsection{Sufficient Statistic as a Function}
\label{subsec:functions}

\begin{figure}
  \includegraphics{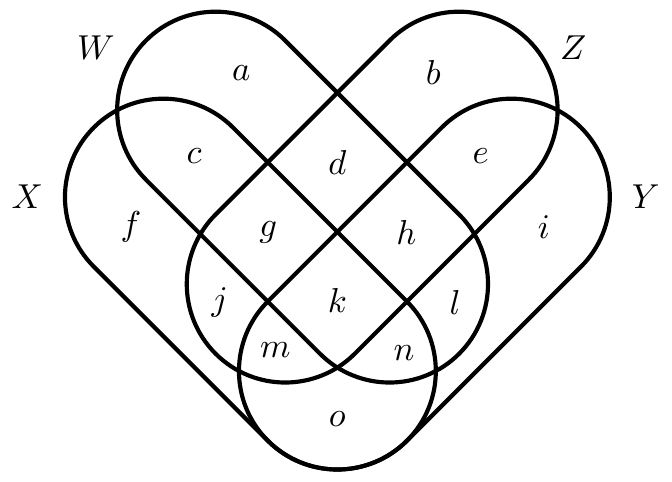}
\caption{Information diagram (I-diagram) for four random variables $X$, $W$,
	$Z$, and $Y$. Each is depicted as a stadium shape and the information atoms
	are obtained by forming all possible intersections. Individual atoms
	are identified with lowercase letters.
  }
\label{fig:XWZY_idiagram}
\end{figure}

Our first step in reducing Fig.\nobreakspace \ref {fig:XWZY_idiagram} is to consider the fact that
$W = X \mss Y$ is a function of $X$.\footnote{By $Y = f(X)$, we mean for
all $x$, $\left|\left\{ y : p(Y = y | X = x) > 0 \right\}\right| = 1$.} Any $W
= f(X)$ if and only if $\H{W \mid X} = 0$~\cite{Shal98a}.
Furthermore, conditional entropies $\H{W
\mid \bullet}$ are never increased by conditioning on additional
variables~\cite{Cove06a}. Since conditional entropies are
nonnegative~\cite{Cove06a}, conditioning $W$ on variables in addition to $X$ can only yield additional zeros.
In terms of the information atoms, the relations:
\begin{align*}
  \H{W \mid X}     &= a + d + h + l = 0 \\
  \H{W \mid X,Y}   &= a + d = 0. \\
  \H{W \mid X,Z}   &= a + l = 0. \\
  \H{W \mid X,Z,Y} &= a = 0,
\end{align*}
imply $a = d = h = l = 0$. A symmetric argument implies that $b = d = g = j = 0$. Each of these zeros is marked with an asterisk in Fig.\nobreakspace \ref {fig:suff_stat_idiagram}.

\subsection{Sufficient Statistic as a Markov Chain}
\label{subsec:markov_chain}

Variables $X$, $V$, and $Y$ form a Markov chain $X - V - Y$ if and only if
$\I{X : Y \mid V} = 0$. Said informally, $V$ statistically shields $X$ and $Y$,
rendering them conditionally independent. Applied to variable $W$ we
find:
\begin{align*}
  \I{X : Y \mid W} & = 0 \\
             m + o & = 0
  ~,
\end{align*}
and similarly for $Z$,
\begin{align*}
  \I{X : Y \mid Z} & = 0 \\
             n + o & = 0
  ~.
\end{align*}
Since $o = \I{X : Y \mid W, Z}$ is a conditional mutual information, $o$ is nonnegative by the standard Shannon inequality~\cite{Cove06a}.

Thus far, $m$ and $n$ are not individually constrained and so could be negative.
However, consider $\I{X : Z \mid W} = j + m$, another conditional mutual information, which is therefore also nonnegative. It is already known that $j=0$, therefore $m$ is nonnegative. Clearly, then, $m$ and $o$ are individually zero.

\begin{figure}
  \includegraphics{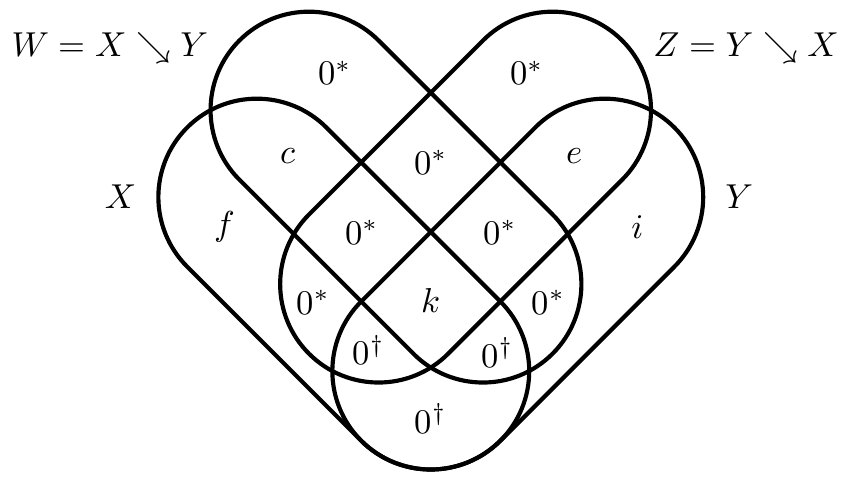}
  \caption{
    I-diagram for sufficient statistics: The vanishing information atoms implied by a sufficient statistic being a function of a random variable are labeled $0^*$. Those vanishing atoms implied by a sufficient statistic forming a Markov chain are marked with $0^\dagger$.
  }
  \label{fig:suff_stat_idiagram}
\end{figure}

Analogously, we find that $n$ is nonnegative and conclude that $n$ and $o$ are individually zero. These vanishing atoms are marked with $0^\dagger$ in the simplified I-diagram in Fig.\nobreakspace \ref {fig:suff_stat_idiagram}.

From this reduced diagram we can easily read that:
\begin{align}
  k &= \I{X:Y} \label{eq:MIs} \\
    &= \I{X:Z} \nonumber \\
    &= \I{W:Y} \nonumber \\
    &= \I{W:Z} \nonumber \\
    &= \I{X:W:Z} \nonumber \\
    &= \I{X:W:Y} \nonumber \\
    &= \I{X:Z:Y} \nonumber \\
    &= \I{W:Z:Y} \nonumber \\
    &= \I{X:W:Z:Y} \nonumber
  ~.
\end{align}

\begin{figure}
  \includegraphics{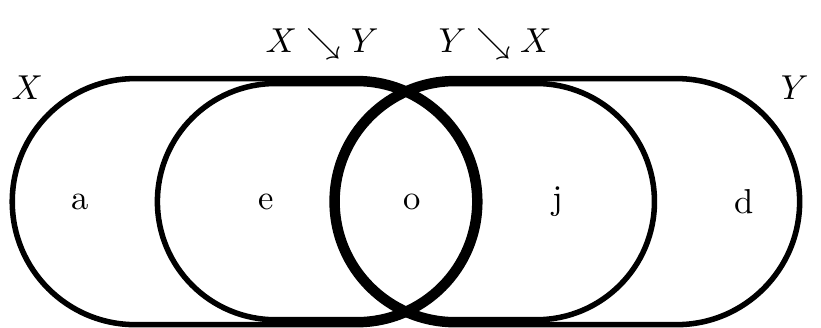}
  \caption{
    Minimal I-diagram containing only nonvanishing atoms in Fig.\nobreakspace \ref {fig:suff_stat_idiagram}.
  }
  \label{fig:reduced_idiagram}
\end{figure}

Furthermore, one can remove the atoms that vanish to arrive at the reduced I-diagram of Fig.\nobreakspace \ref {fig:reduced_idiagram}. It contains only five nonzero atoms.

\section{Stochastic Processes as Channels}
\label{sec:channels}

We find useful application of this result in the analysis of stationary stochastic
processes. Computational mechanics \cite{Crut12a} is an information-theoretic
framework for analyzing structured stochastic processes. There, a process is
considered a channel that communicates its (semi-infinite) past
$\MS{-\infty}{0}$ to its (semi-infinite) future $\MS{0}{\infty}$ through the
present \cite{Crut08a,Crut08b}. (The following suppresses $\infty$ when
indexing.) An important process property---\emph{excess entropy}---is the
mutual information $\EE = \I{\Past:\Future}$ between the past and future. $\EE$
is the amount of uncertainty in the future than can be removed by observing the
past.

At first blush, it is not clear how to proceed in computing a mutual
information between two infinite-dimensional random variables such as this.
The answer lies in the concept of causal states.
Causal states play a central role as the minimal effective states of a
process' channel. The \emph{forward-time causal states}
comprise the minimal amount of information from the past required for
predicting the future.
More precisely, the random variable $\FCS$ is the minimal
sufficient statistic of the past about the future.
 Analogously, the \emph{reverse-time causal states} $\RCS$ embody the minimal sufficient statistic of the future about the past---the states needed for optimally retrodicting the past from the future.

By making the following substitutions: $X \to \Past$, $W \to \FCS$, $Z \to \RCS$, and $Y \to \Future$ in Eq.\nobreakspace \textup {(\ref {eq:MIs})}, we immediately see that the excess entropy (past-future mutual information) has several alternate expressions:
\begin{align}
  \EE & \equiv \I{\Past:\Future} \\
      &= \I{\Past:\RCS} \nonumber \\
      &= \I{\FCS:\Future} \nonumber \\
      &= \I{\FCS:\RCS}
	  ~.
\end{align}
The last identity gives our main result: The excess entropy is the mutual
information between the forward-time and reverse-time causal states. As such,
this provocatively suggests a communication channel between the forward- and
reverse-causal-state processes---a channel that determines the amount
information being transmitted through the present. See also Fig.~1 in
Ref.~\cite{Crut08a}, analogous to Fig.\nobreakspace \ref {fig:reduced_idiagram}.

We can interpret this operationally. Consider a past $\past$, the particular
forward-time causal state $\FCS$ it induces, and an instance $\future$ of the
future following this state. This future analogously induces a reverse-time
causal state $\RCS$. Considering the above channel between forward- and
reverse-time states, the forward state $\FCS$ corresponds to a distribution
over reverse-time causal states $\RCS$. Sampling a state from this distribution
results in a state that gives as much information (retrodictivity) about the
past as the particular reverse state determined by the future.

Continuing, there are a number of related multivariate mutual information~\cite{Jame11a} identities that following directly:
\begin{align*}
\EE   &= \I{\Past:\FCS:\RCS} \\
      &= \I{\Past:\FCS:\Future} \\
      &= \I{\Past:\RCS:\Future} \\
      &= \I{\FCS:\RCS:\Future} \\
      &= \I{\Past:\FCS:\RCS:\Future}
  ~.
\end{align*}

Furthermore, making use of the vanishing information atoms, we find that the following Markov chains exist:
\begin{align*}
  &\Past - \FCS - \RCS - \Future ~, \\
  &\FCS - \Past - \RCS - \Future ~, \\
  &\Past - \FCS - \Future - \RCS ~,~\text{and} \\
  &\FCS - \Past - \Future - \RCS
  ~.
\end{align*}

Causal states are, as noted, \emph{minimal} sufficient statistics. This
minimality is not necessary in the above development. As defined in Ref.
\cite{Shal98a}, a \emph{prescient state} $\PrescientState$ is one for which
$\I{\Past : \Future \mid \PrescientState_0} = 0$ and $\PrescientState$ is a
function of the past. In contrast to the causal states, prescient states need
not be minimal. And so, with little else said, the analogous results follow for
predictive and retrodictive prescient states. For example, we have $\EE =
\I{\PrescientState^-:\PrescientState^+}$.

If we were to lift the restriction that prescient states are functions of the
past (or the future), the resulting forward and reverse
generative~\cite{lohr2009generative} states may interact in their ``gauge''
informations. That is, the atom labeled $d$ in Fig.\nobreakspace \ref {fig:XWZY_idiagram} may be
nonzero; for more on this, see Ref.~\cite{Elli11a}.
The utility of our mutual information identities is then unclear.

The excess entropy, and related information measures, are widely-used diagnostics for complex systems, having been applied to detect the presence of organization in dynamical systems~\cite{Fras86a,Casd91a,Spro03a,Kant06a}, in spin systems~\cite{Crut97a,Erb04a}, in Markov random fields~\cite{Bula09a}, in neurobiological systems~\cite{Tono94a,Bial00a,Marz14e}, in long-memory processes~\cite{Marz15a}, and even in human language~\cite{Ebel94c,Debo08a}.

With these application domains in mind, we should call out the analytical
benefits of using causal states, along the lines analyzed here. The benefits
are particularly apparent in Refs.~\cite{Marz14e,Marz15a}, for example. While
closed-form expressions for excess entropy of finite-state processes have
existed for several years~\cite{Crut08a,Crut08b}, it is only recently that it
has been analyzed for truly complex (infinite-state)
processes~\cite{Marz14e,Marz15a}. In this work, identifying and then framing calculations
around the causal states led to substantial progress. The detailed results here
show why this is true: as sufficient statistics, causal states capture the
essential structural information in a process. Similar benefits should also
accrue when developing empirical estimation and inference algorithms for
related information measures.

\begin{figure}
  \includegraphics{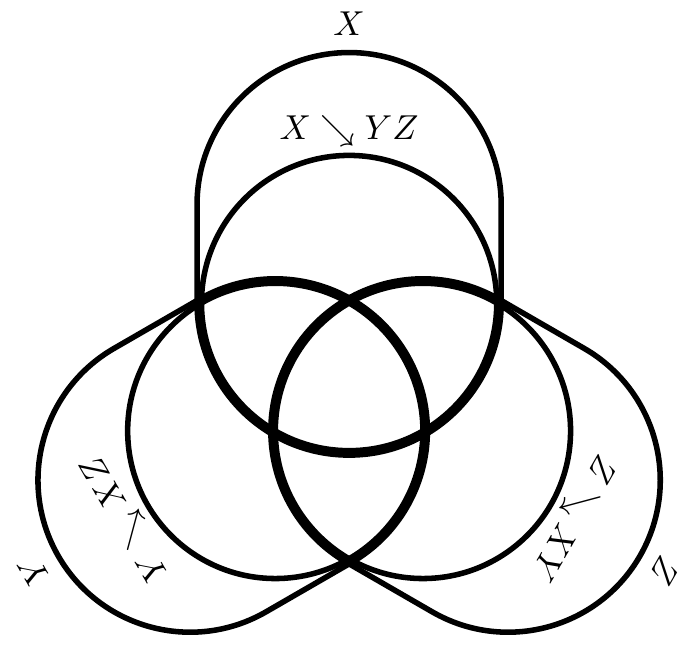}
\caption{Minimal I-diagram involving three variables and their minimal
	sufficient statistics. This differs from a standard $3$-variable I-diagram
	by the addition of three atoms: $\H{X \mid X \mss YZ}$, $\H{Y \mid Y \mss
	XZ}$, and $\H{Z \mid Z \mss XY}$.
  }
  \label{fig:idiagram_three}
\end{figure}

\section{Multivariate Extensions}
\label{sec:multivariate_extensions}

The results can be extended to multivariate systems as well as
to alternative measures of shared information. Consider a system of three
variables $X$, $Y$, and $Z$. The I-diagram of interest involves six
variables: $X$, $Y$, $Z$, and their sufficient statistics about the other
variables: $X \mss YZ$, $Y \mss XZ$, and $Z \mss XY$. This I-diagram contains
$2^{6} - 1 = 63$ atoms. It can be substantially simplified along the
lines of the previous section. First, note that if $A$, $B$, $C$, and $D$
form the Markov chain $A - B - CD$, then we also have the chains $A - B - C$
and $A - B - D$. Second, recall our primary result that $\I{X:Y} = \I{X \mss
YZ : Y \mss XZ}$ and note there are similar relations for the pairs $(X, Z)$ and
$(Y, Z$). Combining these two observations and the methods employed in
Section\nobreakspace \ref {sec:sufficient_statistics} allows one to determine that $53$
atoms are identically $0$. This reduction results in the I-diagram of
Fig.\nobreakspace \ref {fig:idiagram_three}.

Remarkably, the structure of this reduced I-diagram allows us to immediately
conclude that the \emph{total correlation}
$\T{X:Y:Z}$~\cite{watanabe1960information}, \emph{dual total correlation}
$\B{X:Y:Z}$~\cite{sun1975linear}, \emph{co-information}
$\I{X:Y:Z}$~\cite{Bell03a,mcgill1954multivariate}, \emph{CAEKL mutual
information}
$\J{X:Y:Z}$~\cite{chan2015multivariate}, and any other multivariate
generalization of the mutual information remains unchanged under substitution
of sufficient statistics. That is:
\begin{align*}
  \T{X:Y:Z} &= \T{X \mss YZ : Y \mss XZ : Z \mss XY} ~,\\
  \B{X:Y:Z} &= \B{X \mss YZ : Y \mss XZ : Z \mss XY} ~,\\
  \I{X:Y:Z} &= \I{X \mss YZ : Y \mss XZ : Z \mss XY} ~, ~\text{and}\\
  \J{X:Y:Z} &= \J{X \mss YZ : Y \mss XZ : Z \mss XY}
  ~.
\end{align*}

We conjecture that this behavior holds for any number of variables. That is,
replacing each variable by its sufficient statistic about the others does not
perturb the informational interactions among the variables. Nor does it induce
any additional interactions among the sufficient statistics. And so, any
multivariate mutual information will be invariant. We further conjecture that
this is true of any common information, such as the \emph{G{\'a}cs-K{\"o}rner
common information}~\cite{Gacs73a,tyagi2011function}, the \emph{Wyner common
information}~\cite{wyner1975common,liu2010common}, and the \emph{exact common
information}~\cite{kumar2014exact}.

\section{Concluding Remarks}
\label{sec:conclusion}

We demonstrated that it is proper to replace each variable with a sufficient
statistic about its other variables without altering information-theoretic
interactions among the variables. This is a great asset in many types of
analysis and provides a principled method of performing lossless dimensionality
reduction. As an important specific application, we demonstrated how the causal
states of computational mechanics allow for the efficient computation of the
excess entropy.

Our proof method centered around the use of an I-diagram and its atoms. Steps
in our proof, such as identifying that the atom labeled $m$ is nonnegative via
its containment in $\I{X:Z \mid W}$, are greatly aided by this graphical tool.
Despite this, we believe that a superior proof of these results exists---a
proof that does not depend on demonstrating atom-by-atom that all but a select
few are zero. Such a proof would, hopefully, apply generically and directly to
an $N$-variable system, hold for the menagerie of multivariate generalizations
of the mutual information, and perhaps apply even to the common informations.

\section*{Acknowledgments}
\label{sec:acknowledgments}

We thank Dowman P. Varn for helpful conversations. This material is based upon work supported by, or in part by, the John Templeton Foundation grant 52095, the Foundational Questions Institute grant FQXi-RFP-1609, and the U. S. Army Research Laboratory and the U. S. Army Research Office under contracts W911NF-13-1-0390 and W911NF-13-1-0340.

\end{document}